\documentclass{./aa}
\usepackage{graphics}

\begin{document}

   \title{Astrometry of the stellar image of U Her amplified by the
   circumstellar 22~GHz water
   masers}
   \titlerunning{Astrometry of the U Her water masers}


   \author{W.H.T. Vlemmings\inst{1}\and
        H.J. van Langevelde\inst{2}\and
        P.J. Diamond\inst{3}\
          }

   \offprints{WV (vlemming@strw.leidenuniv.nl)}

   \institute{Sterrewacht Leiden, Postbus 9513, 2300 RA Leiden, 
              the Netherlands
        \and
              Joint Institute for VLBI in Europe, Postbus 2, 
                7990~AA Dwingeloo
        \and
         Jodrell Bank Observatory, University of Manchester, Macclesfield,
                    Cheshire, SK11 9DL, England     
        }

   \date{Received ; accepted }


\abstract{ The 22~GHz H$_2$O masers in the circumstellar envelope of
the Mira variable star U~Her have been observed with MERLIN using a
phase referencing technique to determine accurate astrometric
positions. The positions were compared with the optical positions
obtained with the Hipparcos satellite to an accuracy of $18$~mas. The
absolute radio position of the brightest H$_2$O maser spot is found to
match the optical position, indicating that this spot is the stellar
image amplified by the maser screen in front of it.  The occurrence of
an amplified image in the 22 GHz maser can be used to accurately
determine the positions of the H$_2$O with respect to the star as well
as with respect to the SiO and OH masers. Our observations seem
to indicate that the star is not in the centre of the distribution of
maser spots, which has been interpreted as a ring.  }

   \maketitle

\section{Introduction}

The circumstellar envelopes (CSEs) around late-type stars contain
several maser species that are excellent probes of the dynamics of the
outflowing material. Astrometric observations are an important tool to
study the locations and motions of the circumstellar masers with
respect to the central star. This understanding is essential to reach
the main goal of maser astrometry, which is to determine the distances
to heavily enshrouded stars, that are too faint for their parallax to
be determined directly.

\subsection{Circumstellar H$_2$O masers}

Interferometric observations of the 22 GHz H$_2$O masers in the CSEs
of Mira-variable stars with MERLIN, the VLA and Very Long Baseline
Interferometry (VLBI) indicate that the H$_2$O masers are found up to
a few hundred AU from the star (e.g. Lane et al. 1987). This is
generally inside the OH maser shell, which is located at up to several
1000 AU.  The H$_2$O masers often show an a-spherical distribution,
and the size of the maser region is thought to increase with mass-loss
rate (Cooke \& Elitzur 1985). The maser is expected to be pumped due
to collisions (Neufeld \& Melnick 1991). The 22 GHz maser can then be
easily excited in the inner parts of the CSE at temperatures of 400 to
1000 K and H$_2$ number densities of $10^9$ cm$^{-3}$. In this region
the outflow is still being accelerated. Therefore, as shown by
Rosen at al. (1978) the velocity coherent paths through the masing
medium are of approximately equal length in both the radial direction
as well as the direction tangential to the star, and thus the H$_2$O
maser beaming is expected to be both radial and tangential. The radial
beaming results in the maser occurring in front of the star,
tangential beaming would display a ring within a narrow velocity range
close to the stellar velocity (e.g. Reid \& Menten 1990).

H$_2$O masers are significantly more variable than their OH cousins.
They exhibit strong variability in intensity, which seems to indicate
that the masers are at least partially unsaturated, since an
unsaturated maser is strongly influenced by changes in the local
conditions. An analysis of the H$_2$O maser line-widths and
line-shapes also indicates that the masers are not completely
saturated (e.g. Vlemmings et al. 2002).

 Whereas semi-regular stars are observed to have H$_2$O maser spectra
that can change shape rapidly, Mira-variable stars typically show no large
profile changes over several years. However, the individual features
can still show significant changes in intensity (Engels et al. 1988).

\subsection{Amplified Stellar Image}

Interferometric observations of OH masers have revealed that the most
compact features were only found at the blue-shifted side of the
spectrum (Norris et al. 1984). It was argued that this was due to
amplification of the continuum maser emission from the underlying star
by the maser screen in front of it. This is called the {\it
Amplified Stellar Image Theory}. Amplification of the stellar emission
results in a high brightness maser spot at the most blue-shifted
side of the OH maser spectrum. This spot should coincide at the
different OH maser transitions, and is expected to be persistent over
a long period of time. Several observations have confirmed this
hypothesis (e.g. Sivagnanam et al. 1990). 

According to the amplified stellar image theory, the compact, most
blue-shifted, spot is necessarily fixed to the stellar position.
Therefore, high resolution astrometric observations of this spot can
be used to determine the stellar trajectory. This hypothesis has been
tested for the Mira-variable star U~Her by van Langevelde et
al. (2000, hereafter vL00). The {\it absolute} positions of the OH maser
spots were determined with respect to the radio-reference frame using
extra-galactic phase reference sources. The positions were compared
with the optical Hipparcos positions with unprecedented
accuracy. It was shown that the most blue-shifted spot was indeed
located in front of the stellar radio-photosphere. The size of this
spot was found to be $\approx 20$~mas, which is comparable with the
expected size of the radio-photosphere, which is thought to be twice
the size of the star, as proposed by Reid \& Menten (1997).

Although the H$_2$O masers generally show a great number of spots over
an area of several hundred mas, it has been argued that in some cases
one of the H$_2$O maser spots corresponds to the stellar image (Reid
\& Menten 1990; Marvel 1996; Colomer et al. 2000). However, because
the distribution of the H$_2$O masers is considerably less spherical
than that of the OH masers, it is not straightforward to assume that
the stellar image underlies the most blue-shifted spot. Also, because
the maser brightness depends strongly on local effects such as density
or pumping inhomogeneities, several bright spots can be observed and
an H$_2$O maser stellar image could be less conspicuous or persistent
than the OH stellar image.

 Here we present phase referencing observations of the H$_2$O masers
around U~Her used to determine accurate maser spot positions. These
have been compared with the Hipparcos optical position and the
positions obtained for the OH masers in vL00.

\section{Observations}

MERLIN was used to determine the positions of the H$_2$O
masers in the CSE of U~Her with respect to 3 extragalactic reference
sources. From the observations, which were performed on May 20 2001,
we were able to determine accurate astrometric positions.  The
observations required a total of 8 hours on the target source and the
calibrator and reference sources. As a result the beam size was
$\approx 30 \times 25$ mas.

One of the main problems of using the phase referencing method to
determine accurate positions at 22 GHz is the lack of bright reference
sources at that frequency. For this project 3 reference sources were
selected from the VLBI calibrator catalog (Beasley et al.,
2002). These were J1628+214 (now J1630+231, $\approx 80$~mJy, at
$2.8^\circ$ from U~Her), J1635+1831 ($\approx 50$~mJy, at $2.4^\circ$)
and J1619+2247 ($\approx 200$~mJy, at $4.2^\circ$). They were selected
to have a flat spectrum, making it possible to detect them at 22~GHz
with a 2 minute integration time. Our phase reference observation
cycle consisted of 2 minutes per reference source and 4 minutes on the
target source.  The H$_2$O masers were observed in a $4$ MHz band with
128 spectral channels, providing a velocity resolution of $0.42$
km/s. The continuum reference sources were observed in a $16$ MHz band
with 16 channel, giving a $13.5$ km/s resolution. In the 2 minute
integration time on the reference source we typically reached a
SNR$\approx 5$.

Reference source J1628+214 was also used in vL00 to determine the
positions of the 1667 MHz OH masers around U~Her. We can thus
determine the positions of the H$_2$O masers with respect to the OH
masers. The positional accuracy of the reference sources with respect
to the radio reference frame is $\approx 3$ mas. However, as the
positions were determined at $2.4$ and $8.4$~GHz the position at the
different frequencies can be different at the sub-milliarcsecond
level, as shown in e.g. Fey et al.(1997).

After initial flux calibration with the MERLIN software, the data were
processed in AIPS. Because the reference sources at 22~GHz are weak,
we performed both a normal and a reverse phase referencing scheme.
In the reverse scheme, we were able to determine accurate
positions of both J1619+2247 and J1628+214 with respect to the
brightest H$_2$O maser feature of U~Her. As a consistency check, we
managed to get a good phase connection from J1628+214, the closest
reference source, to the U~Her H$_2$O masers with the normal phase
referencing scheme. J1635+1831 was too faint for us to obtain a good
phase connection.

The accuracy of the phase referencing model depends on the MERLIN
correlator model. The two largest uncertainties in the model are the
telescope positions (accurate to $\sim 5$~cm) and tropospheric
effects.  Under normal circumstances these combine to produce an error
in absolute position measurements of up to $\sim 10$ ~mas for a
target-reference source separation of $\sim 3^\circ$ (e.g. Kovalevsky
et al 1997).

\section{Results}

\begin{figure} 
   \resizebox{\hsize}{!}{\includegraphics{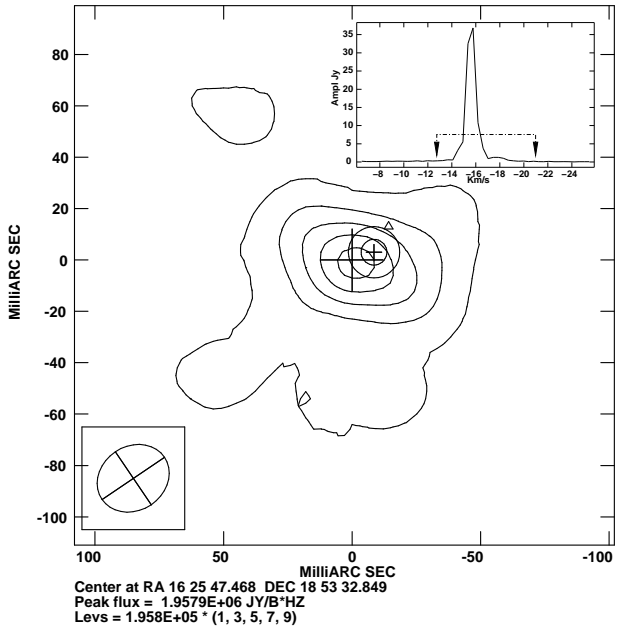}} \hfill
   \caption{The location of the U~Her H$_2$O masers with respect to
   the stellar positions determined from the Hipparcos observations,
   as determined from phase referencing to J1628+214. The star is
   denoted by two circles indicating the star itself and the stellar
   radio photo-sphere. The large cross indicates the position of the
   H$_2$O maser, the error bars are due to the positional fitting. The
   errors on the stellar position (small cross) are due to the link to
   the radio-reference frame and due to the errors in the proper
   motion used to transpose the optical position. The triangle denotes
   the stellar position when using the phase referencing results on
   the J1619+2247 reference source. The inlay is the averaged
   cross-power spectrum of the H$_2$O maser emission. The masers were
   mapped over the velocity range indicated in the spectrum.}
   \label{fig1}
\end{figure}

 At the time of our observations the spectrum of the 22 GHz H$_2$O
masers around U~Her did not show significant structure, as only a few
features were detected. The averaged cross power spectrum is shown in
an inlay in Fig.~\ref{fig1}. We find that the shape of the spectrum is
similar to previous observations, performed with MERLIN, the Very
Large Array (VLA) and with the 100-m radio telescope in Effelsberg
(Baines et al. in preparation; Yates \& Cohen 1994; Colomer et
al. 2000; Engels et al. 1988). In our spectrum we find that the
strongest feature is located at $-15.7$ km/s, and this is the feature
which was used to determine the phase solutions. The stellar velocity
of U~Her is $-14.5 \pm 0.5$ km/s, which was determined from OH and SiO
maser observations (Chapman et al. 1994). As the U~Her H$_2$O maser
emission is located between $-13$ and $-20$ km/s, the strongest
feature is not the most blue-shifted feature.

After phase referencing and determining the position of J1628+214 with
respect to the brightest H$_2$O maser feature of U~Her, we find that
the position of J1628+214 is shifted by $+83$ mas in right ascension
and $+15$ mas in declination with respect to the positions in vL00
($(\alpha,\delta)_{2000}=16^h30^m11^s.23117$,$+21\degr31\arcmin34\arcsec.3144$).
A similar process for J1619+2247 results in a position shift of $+76$
mas in right ascension and $+25$~mas in declination with respect to
the position in the VLBA calibrator list.  As a consistency check we
have also determined the position shift of the brightest maser spot
after phase referencing with respect to J1628+214. We find that the
maser spot is shifted $-83$ and $-11$ mas in right ascension and
declination respectively with respect to the {\it a priori} assumed
target coordinates.

The formal uncertainty in fitting a Gaussian profile to the reference
source or maser spot is a fraction of the beam width, and depends on
the SNR of the image. For the reference sources the formal position errors
are of the order of $5$~mas in each coordinate, for the U~Her maser
spot the errors are $\approx 1$~mas. The best phase connection was
made when phase referencing J1628+214 to the brightest maser spot, so
we assume that the positions as determined with J1628+214 are the most
reliable. The actual phase referencing errors can be estimated from
the difference between the position of the brightest maser spot with
respect to the two reference sources. From this, we conclude that our
positions are accurate to within $10$~mas, which is in agreement with
the estimated errors due to the correlator model that are described
above. For the brightest H$_2$O maser spot around U~Her we then find a
position of
$(\alpha,\delta)_{2000}=16^h25^m47^s.468$,$+18\degr53\arcmin32\arcsec.849$
at the time of our observations.

 To compare this position to the stellar position of U~Her we have
extrapolated the optical position found by the Hipparcos satellite at
J1991.25 to our epoch of observation. We have used the proper motion
and parallax determined by monitoring the position of the most
blue-shifted OH maser spot, which was shown to be the stellar
image. The first fit was performed in vL00 for 6 epochs of
observations, a fit including additional epochs was presented in
Vlemmings et al. (2000). As described in vL00, the OH maser proper
motion is entirely consistent with the Hipparcos proper motion. The
error in the transposed position is dominated by the error in proper
motion. At our epoch of observation this error is $\approx 6$ mas in
each coordinate. Combined with the errors on the parallax and our
position errors, we have been able to compare the radio and optical
position with $\approx 18$ mas accuracy.  Fig.~\ref{fig1} shows a map
of the H$_2$O maser features, covering the velocity range indicated in
the spectrum, including the position of the star. Circles indicate the
size of the star and the radio-photosphere. The size of the
radio-photosphere can be estimated from SiO maser observations by
Diamond et al. (1994). Their observations provide an upper limit of
$\approx 20$~mas if, as proposed by Reid \& Menten (1997), the
radio-photosphere extends to the edge of the SiO masing region. The
triangle denotes the stellar position as determined when using the
maser positions with J1635+1831 as the reference source.

 Although the size of the brightest H$_2$O maser spot ($\approx
50$~mas) is larger than the expected size of the stellar
radio-photosphere, most likely due to the blending of several weaker
maser features, the position of peak intensity matches the predicted
location of the star within the errors. This indicates that the H$_2$O
maser spot also coincides with the most blue-shifted OH maser spot
which has been shown to be the amplified stellar image. Thus, also the
brightest H$_2$O maser spot seems to be emission from the stellar
radio-photosphere amplified by the maser medium at the line of sight.

\section{Discussion}

 The observations of an amplified stellar image in the H$_2$O masers
of U~Her indicate that at least for the brightest spot at this epoch
the maser beaming is radial. Reid \& Menten (1997) have detected
22~GHz continuum emission from a small sample of Mira stars, finding
the typical stellar brightness temperature to be $T_* =
1600$~K. Compared to their estimate of the maser excitation
temperature ($\approx 10$~K), this is strong enough to influence the
H$_2$O maser medium and produce a stellar image, as the increased seed
radiation from the star will cause the radial maser beam to be
brighter. Since the H$_2$O masers are found to be mostly unsaturated,
slight changes in density, pumping and velocity structure have a
strong effect on the maser and the relative strength of the maser
features and the amplified stellar image may be less dominant than the
OH, as is demonstrated by the detection by Reid \& Menten (1990) of an
H$_2$O maser feature at the stellar position of W~Hya which was
several orders of magnitude weaker than the strong feature observed
here.

The H$_2$O masers around U~Her have been observed before with MERLIN,
the VLA and the VLBA. VLA observations by Colomer et al. (2000), and
MERLIN observations by Bains et al. (in preparation) show an
incomplete ring structure with a scale of $150-200$~mas. The brightest
maser spot seen in our MERLIN observations corresponds in velocity
with the masers on the edge of the ring structure. So somewhat
surprisingly, our astrometric results indicate that the star is not in
the center of this ring.

The maser
spots detected with high resolution VLBA observations do not show any
indication of circular structure (Vlemmings et al. 2002). They have a
linear extent of $\approx 60-70$~mas and they most likely correspond
to the brightest VLA and MERLIN features. 

\section{Conclusions}

 Our observations have shown that the circumstellar H$_2$O maser can
amplify the stellar image and produce a strong stellar image, a
phenomenon previously detected at the OH maser transitions.  Although
this effect is not necessarily strong in all H$_2$O masers, it can be
very valuable for astrometric purposes. Accurate astrometry of the
stellar image can be used to determine the location of the various
maser species in the CSE with respect to each other and the
star. Simultaneous, high resolution observations significantly improve
our understanding of the kinematics in the CSEs. Using the stellar
image in H$_2$O masers, it will also be possible to determine the
stellar trajectory and distance with a higher accuracy than with OH
masers. However, because of the high variability of H$_2$O masers
additional monitoring will have to be performed to show if the stellar
image is persistent enough for a long term monitoring campaign.

{\it Acknowledgments:} This project is supported by NWO grant
 614-21-007. MERLIN is a National Facility operated by the University
 of Manchester at Jodrell Bank Observatory on behalf of PPARC. We also
 thank the referee Mark Reid for valuable input.

\end{document}